\documentclass{aa}  
\usepackage{graphicx}
\usepackage{txfonts}
\usepackage{lipsum}
\usepackage[normalem]{ulem}
\usepackage{color}

\begin{document}

\newcommand{\thecode}{FLATW'RM}

   \title{Finding flares in {\em Kepler} data using machine-learning tools}

%   \subtitle{The subtitle}

   \author{Kriszti\'an Vida
          \inst{1}
          \and
          Rachael M. Roettenbacher\inst{2}
          }
\authorrunning{K. Vida \& R. M. Roettenbacher}
   \institute{
        Konkoly Observatory, MTA CSFK, H-1121 Budapest, Konkoly Thege M. \'ut 15-17, Hungary\\
        \email{vidakris@konkoly.hu}
   \and
        Department of Astronomy, Stockholm University, SE-106 91 Stockholm, Sweden
     }

   \date{Received March 1, 2018; accepted March 1, 2018}

% \abstract{}{}{}{}{} 
% 5 {} token are mandatory
 
  \abstract
   {
   Archives of long photometric surveys, such as the {\em Kepler} database, are a great basis for studying flares. However, identifying the flares is a complex task; it is easily done in the case of single-target observations by visual inspection, but is nearly impossible for several year-long time series for several thousand targets. Although automated methods for this task exist, several problems are difficult (or impossible) to overcome with traditional fitting and analysis approaches. 
   }
   {
   We introduce a code for identifying and analyzing flares based on machine-learning methods, which are intrinsically adept at handling such data sets.
   }
   {
   We used the RANSAC (RANdom SAmple Consensus) algorithm to model light curves, as it yields robust fits even in case of several outliers, such as flares. The light curves were divided into search windows, approximately on the order of the stellar rotation period. This search window was shifted over the data set, and a voting system was used to keep false positives to a minimum: only those flare candidate points were kept that were identified as a flare in several windows.
   }
   {The code was tested on short-cadence {\em K2} observations of TRAPPIST-1 and on long-cadence {\em Kepler} data of KIC 1722506. The detected flare events and flare energies are consistent with earlier results from manual inspections.}
   {}

   \keywords{
   Methods: data analysis --
   Techniques: photometric --
   Stars: activity --
   Stars: flare --
   Stars: late-type --
   Stars: low-mass 
   }

   \maketitle
%
%-------------------------------------------------------------------

%%%%%%%%%%%%%%%%%%%%%%%%%%%%%%%%%%%%%%%%%%%%%%%%%%%%%
\section{Introduction}  % I N T R O D U C T I O N   %
%%%%%%%%%%%%%%%%%%%%%%%%%%%%%%%%%%%%%%%%%%%%%%%%%%%%%

Flares are energetic eruptions that occur as a result of magnetic field line reconnection. These events can be found in almost all types of main-sequence stars, including hot and cool stars \citep{2016ApJ...831....9S,2013ApJS..209....5S}; but flares are most numerous in low-mass, late-type M dwarfs \citep{2011AJ....141...50W, v374peg}.

These energetic events have received increased interest since the advent of exoplanet research, as flares can have strong, deleterious effects on orbiting planets  \citep{2007AsBio...7..167K,2008SSRv..139..437Y}. Flares can also continuously transform exoplanetary atmospheres, which is disadvantageous for hosting life (see \citealt{trappist1,trappist1b} and references therein).  

Currently used definitions of the habitable zone are based only on the stellar irradiation and the distance of the planet from the host star.
As flares can have strong effects on planetary environments, these definitions will likely need to be  revised  
for a more accurate definition of habitability in order to include the effects of stellar activity. To do this, and to better understand stellar magnetism itself, it is essential to characterize flares: events need to be properly identified, and  their strength and frequency need to be determined. 

The data from the {\em Kepler} satellite proved to be a great resource for stellar activity research because they provide an almost continuous data set of unprecedented precision over four years from about 160 000 targets, which include thousands of active stars \citep[e.g.,][]{2010ApJ...713L.155B}.  
Studies have been performed to understand stellar activity of individual stars \citep[e.g.,][]{2013ApJ...767...60R} and of classes of stars \citep[e.g.,][]{2014ApJS..211...24M,appaloosa}.  
%Several studies have been performed to understand the activity of individual stars \citep[e.g.,][]{2013ApJ...767...60R} and large subsets of the catalog\LEt{please check, this is somewhat inadvertedly
%funny: "... studies have been performed to understand... large
%subsets of the catalog" - is this what you wished to say?} \citep[e.g.,][]{2014ApJS..211...24M,appaloosa}.    

Significant strides were made to detect the flares that are contained in the {\em Kepler} archive by \citet{appaloosa}, who identified events in the light curves by detecting the shape of a flare.  However, this method can misidentify other astrophysical phenomena as flares (e.g., KIC 1572802, an RR Lyrae star).  
Of course, there is no single perfect way to accurately detect and classify all flares: the diversity of observations (e.g., short- and long-cadence {\em Kepler} data or ground-based observations) and of the events themselves (the flare length and complexity due to multiple nearly simultaneous flaring events can result in several light-curve shapes) make the automated search for these eruptions a difficult task.  A manual identification of flares is also impossible in practice for a large number of observations,
as in the {\em Kepler} archive.

In this paper, we present an algorithm that is based on machine-learning,
with which we identify flares in light curves\footnote{
The code is available at \url{https://github.com/vidakris/flatwrm/}
}, and we present our application to the flaring, planet-hosting star TRAPPIST-1, a popular target for habitability studies,
%to the planet-hosting star TRAPPIST-1, a popular target for \textcolor[rgb]{0.988235,0.501961,0.0313726}{habitability studies known to flare }\LEt{again, please check - this reads
%now that the studies themselves flare. I assume you mean something
l%ike "for studies of habitability in stars that show flares",
%or similar}\citep[e.g.,][]{trappist1}, 
and KIC 1722506, a rotationally variable star \citep[e.g.,][]{2011A&A...529A..89D}.  

%%%%%%%%%%%%%%%%%%%%%%%%%%%%%%%%%%%%%%%%%%%%%%%%%%%%%%%%%%%%%%%%%%%
\section{Flare-finding algorithm}     % T H E   A L G O R I T H M %
%%%%%%%%%%%%%%%%%%%%%%%%%%%%%%%%%%%%%%%%%%%%%%%%%%%%%%%%%%%%%%%%%%%
\subsection{Determining the stellar rotation period}
The first step of our FLAre deTection With Ransac Method (\thecode{}) algorithm 
is to determine the rotation period of the light curve, which \thecode{} accomplishes by taking the photometric modulation of spotted active stars into account.  The starspots, which are
dark regions of suppressed convection of cool, active stars, rotate across the stellar surface in and out of view of the observer, causing periodic modulations to the light curve.  Starspots are
often longer-lived than sunspots, allowing for detection during multiple rotations. This provides a reliable approximation for the stellar rotation period.  

Light-curve sections of approximately the length of the stellar rotation period (typically, on the order of days) are expected
to be easily described by a relatively low-degree polynomial (as opposed to sections covering several rotations), and their lengths are longer than the timescale of a flare event (typically, on the order of hours).  The light-curve sections are specifically defined such that a flare could be easily spotted in a data set by eye. The period search in \thecode{} is made using \verb+LombScargleFast+, the Lomb--Scargle periodogram implementation in \verb+gatspy+\footnote{\url{http://www.astroml.org/gatspy/}}. For further analysis, these light-curve windows (with a length of $1.5\times P$ by default, where $P$ is the  period found with the Lomb--Scargle method, which is generally assumed to be the rotation period) are used.  
Each light-curve window must also be standardized: it is a common requirement for many machine-learning estimators that individual features should look more or less like normally distributed data (Gaussian with zero mean and unit variance, see, e.g., \citealt{scikit-learn,scikit-book}). 
For our purposes, it is enough to transform only the time axis by removing its mean and scaling the light curve  by its standard deviation, since the scaling of the brightness variation is just a multiplicative factor in the coefficients of the polynomial used to fit the light-curve window.

\subsection{Determining the order of the polynomial fit}
In machine-learning, one major problem is determining the complexity of the model used to fit the data in order to avoid under- or overfitting \citep{scikit-book}. In the case of underfitting, the model does not describe the data well, while in the case of overfitting, a too-complex model is used that tries to fit too many data points individually. While it might fit a training data set well, this will describe test data and future measurements poorly. This problem is generally solved by a cross-validation method. An example of a basic approach is $k$-fold cross-validation, where the training data are split into $k$ smaller sets (so-called folds), and the model is trained on $k-1$ folds of the data. The remaining part of the data is used for validating the model. The sets are usually created by selecting random samples of the initial data set, but this is not very useful with time series.
In these cases, sets of consecutive data points are therefore
used \citep{scikit-learn}. Lengthy data sets, such as {\em Kepler} light curves, make it impractical to use all the available data for cross-validation, or to find the optimal polynomial order for each segment. Therefore, we selected a sample (five, by default) of light-curve windows from the observations, and performed a grid search of fit parameters on them to select the best model to describe the given data based on the median absolute error regression loss. In most cases, a polynomial up to $\approx$10th degree is sufficient to fit rotational modulation of the data in the window.

\subsection{Outlier detection and selection of flare candidates }

\begin{figure}
\begin{center}
\includegraphics[width=0.49\textwidth]{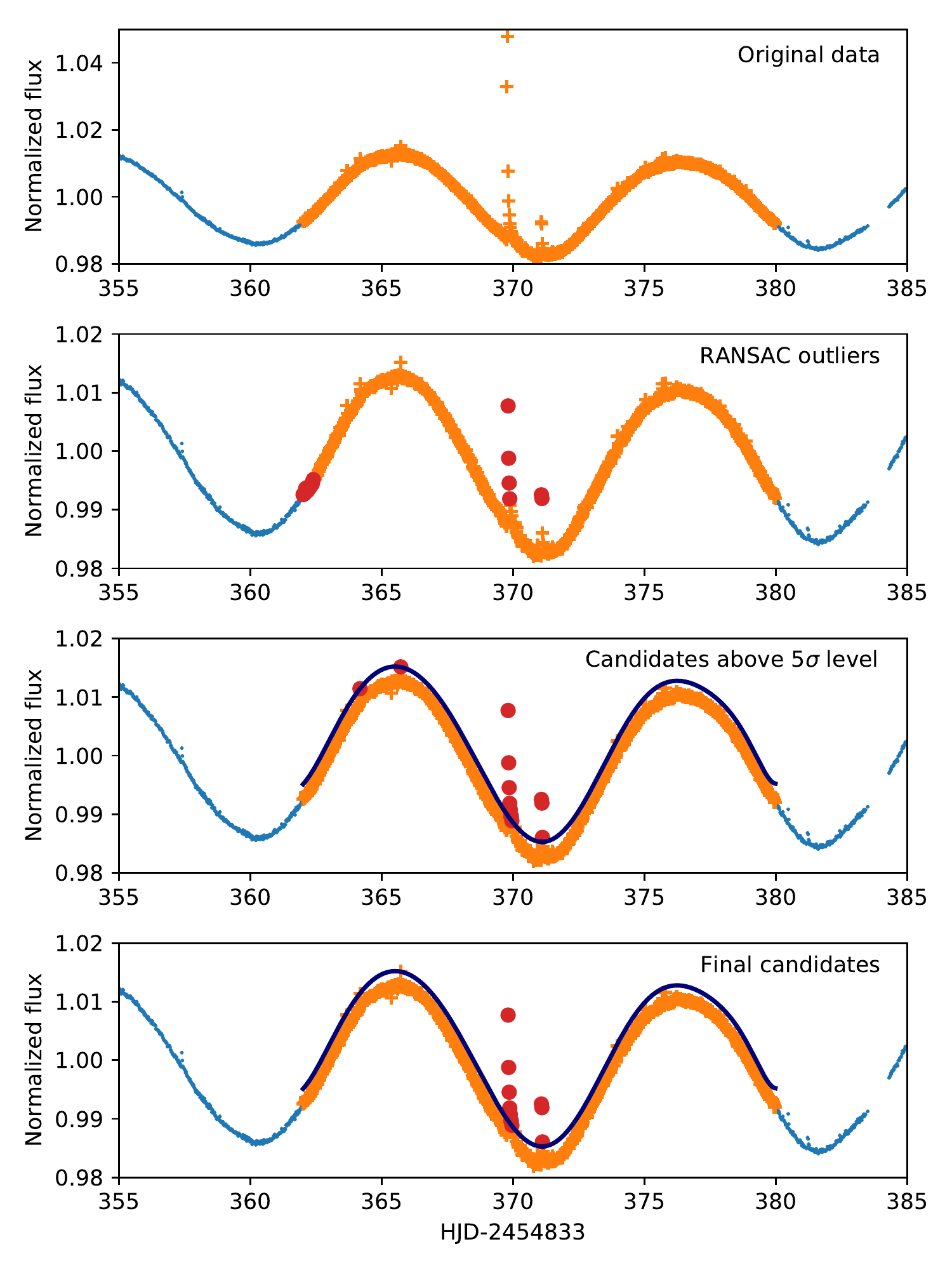}
\caption{Demonstration of the algorithm on a light-curve section of KIC 1722506. The top plot shows the original light-curve section. The second plot shows the outlier points found by RANSAC, marked with red dots. The third plot shows the $5\sigma$ detection level from the fit (continuous line) and the flare candidate points. In the bottom plot, the final flare candidates are shown, which have more than a given number (three, in this case) of consecutive data points. These points will get a vote for this light-curve section, indicating that the feature likely is a flare.    }
\label{fig:test}
\end{center}
\end{figure}

To model light curves, we used the RANdom SAmple Consensus (RANSAC) algorithm, as it is designed to give a robust fit to data with several outliers \citep{ransac}. RANSAC is an iterative method that assumes that the data consist of inlier and outlier points (generally noise, but also flare events, in this case). The algorithm works as follows: first, a sample random subset is generated from the input data set, which is fit by the model. Then, the algorithm checks which elements of the original data set fit this model based on the residuals. The points that fit the model are considered inliers for the given iteration. These steps are iterated either a maximum number of given times or until one of the stop criteria is met (this can be a given number of inlier points or a stop score by a given metrics). The final model is based on all inlier samples (also called a consensus set) of the previously determined best model. An example of the outlier selection by the algorithm is shown in the second plot of Fig. \ref{fig:test}.

\begin{figure}
\begin{center}
\includegraphics[width=0.49\textwidth]{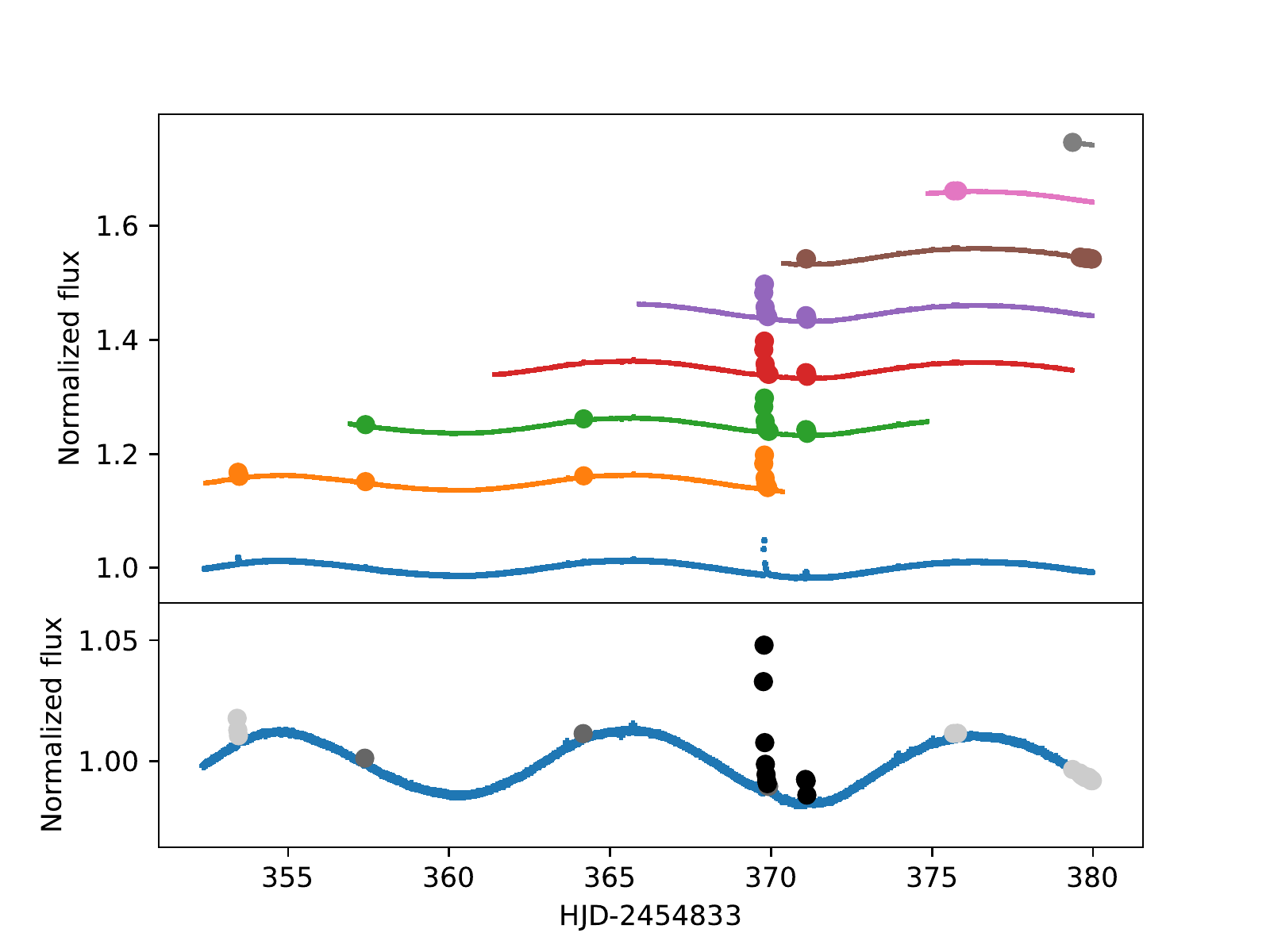}
\caption{Demonstration of the voting algorithm on a light-curve section of KIC 1722506. The top plot shows the original light curve (in blue, near a normalized flux of 1), and each light-curve segment in individual windows tested by \thecode{} (each a different color), and the flare candidates are marked for the given run with circles. In the bottom plot, the candidates with one (light gray), two (medium gray), and at least three (black) votes are plotted. In this setup, the flares plotted in black are kept as final flare candidates.}
\label{fig:voting}
\end{center}
\end{figure}

We found that while RANSAC gives a good fit even for a light-curve section with several flare events, the marked outliers are not reliable enough for searching for flare candidates alone (see Fig. \ref{fig:test}): it sometimes also marks the beginning of the light-curve windows.  Thus we only used the RANSAC estimate of the inlier points for statistics and calculated the standard deviation of the light curve (with the rotational modulation and most of the flare points removed). We considered those points as first-order flare candidates that were above a given detection level (above $3\sigma$ by default). To achieve more robust results, we shifted the search window through the light curve, by default in steps of one-fourth of the light-curve window (see Fig. \ref{fig:voting}, e.g.). In this way, every light-curve point was analyzed multiple times (with the exception of those that are at the beginning or at the end of the data set or large observational gaps), and if a light-curve point is considered as an outlier, that is, a flare point candidate, it received a ``vote''. A light-curve section was considered a flare when it received enough votes ($\ge 3$, by default) in overlapping search windows and had at least a given number (2, by default) of consecutive points. We
note that this last step was performed  only after evaluating each light-curve segment. These criteria are basically the same as those defined by Equations 3 a--d of \cite{2015ApJ...814...35C} and are consistent with adaptations made by \cite{appaloosa}. When running \thecode{} from command line, the user can change the number of flare points needed for a flare, the detection level of the flares, and the full with at half maximum (FWHM) that is used for the analytic fit of the events. Optionally, the rotation period can be given as an input to run the code faster, or to fix it to a chosen value if the rotational modulation is too weak, and the polynomial degree can also be fixed. The number of votes needed for a flare candidate to be kept, the window size (compared to the rotation period), and the step length with which the search window is shifted cannot be changed from command line, but can be easily modified if the flare-finding function is imported to another code.

\subsection{Summary of the selection process}

For clarity, we explicitly state the procedure that \thecode{} steps through in locating the flares in the light curve.

\begin{itemize}
\item A period search is first performed on the input light curve.
\item The input light curve is divided into windows of $1.5 \times P$ that can be effectively fit with polynomials.
\item Each light-curve window is modeled by the RANSAC algorithm to find the best-fitting polynomial. 
\item The data points are designated as inliers or outliers.
\item The light-curve model is subtracted from the light-curve window.
\item The standard deviation of the light curve is determined based only on the inlier data points.
\item The data points that are above the given detection limit receive a vote as a flare candidate for the given window.
\item After each window is analyzed, only those flare candidate data points are kept that have a given number of votes.
\item Events that have at least a given number of selected candidate points are marked as flare events.
\end{itemize}

\subsection{Fitting an analytic model}
As output we considered two options: (1) the beginning and ending times of the flare and the maximum time and  light-curve amplitude for each event, or (2) an analytic model can be fit to the data. In the latter case, the observed light curve is modeled again with the RANSAC algorithm (this time centered on the event) to remove the effect of rotational modulation, leaving only the light-curve changes caused by flares. This data set is then fit by the classical single-peak flare model defined by \cite{2014ApJ...797..122D}. The parameters for this function are the time of the flare peak,
the FWHM (i.e., the timescale of the flare),
and the amplitude of the flare. 
As an initial guess, we took the middle flare data point as the peak time, the amplitude of the selected light-curve data point, and one hour as the timescale of the flare, which fit most of the events (but this can be changed by the user). In the case of weak eruptions near stronger events, however, the fit might be distorted, and the fit could converge to the event with higher amplitude, effectively ignoring the weaker event. In the case of lower sampling, the fits could yield very high peaks (as a consequence of exponential decay) if only the declining phase is measured. 
To estimate flare energies, the integrated intensity (also known as equivalent duration) was also calculated for each event.

%%%%%%%%%%%%%%%%%%%%%%%%%%%%%%%%%%%%%%
\section{Caveats}   % C A V E A T S  %
%%%%%%%%%%%%%%%%%%%%%%%%%%%%%%%%%%%%%%
While this method is an improvement over previous automated flare-detection efforts, we acknowledge that there are still a number of difficulties.  Here, we list the significant caveats for using \thecode{}.

\begin{itemize}
\item In the application of the code on a large number of different light curves (e.g., mixing short- and long-cadence {\em Kepler} data) without adjusting the searching parameters.
\item Long and complex flare events (e.g., those observed by \citealt{dqtau}, where eruptions were likely caused both by accretion and magnetic field reconnection) can cause failed outlier detection if there are not enough inlier points in the search window.
\item Analytic fits to the events can yield unexpected results, especially in the case of long-cadence {\em Kepler} data, where only the exponential decay of the flare is observed (cf. Fig. \ref{fig:samples}).  We emphasize that the analytic fits always
need to be checked or the output from just the light curve should
be used and analytic fitting be ignored in dubious cases.
\item A weak rotation signal may not be properly identified with \thecode{}'s Lomb--Scargle period search and could a yield problematic light-curve search window length.  Additionally, \thecode{} could fail to find a rotation period.  In these cases, the rotation period (or a reasonable size for the light-curve window) should be given as an input to the code.

\end{itemize}

\begin{figure*}
\begin{center}
\includegraphics[width=0.7\textwidth]{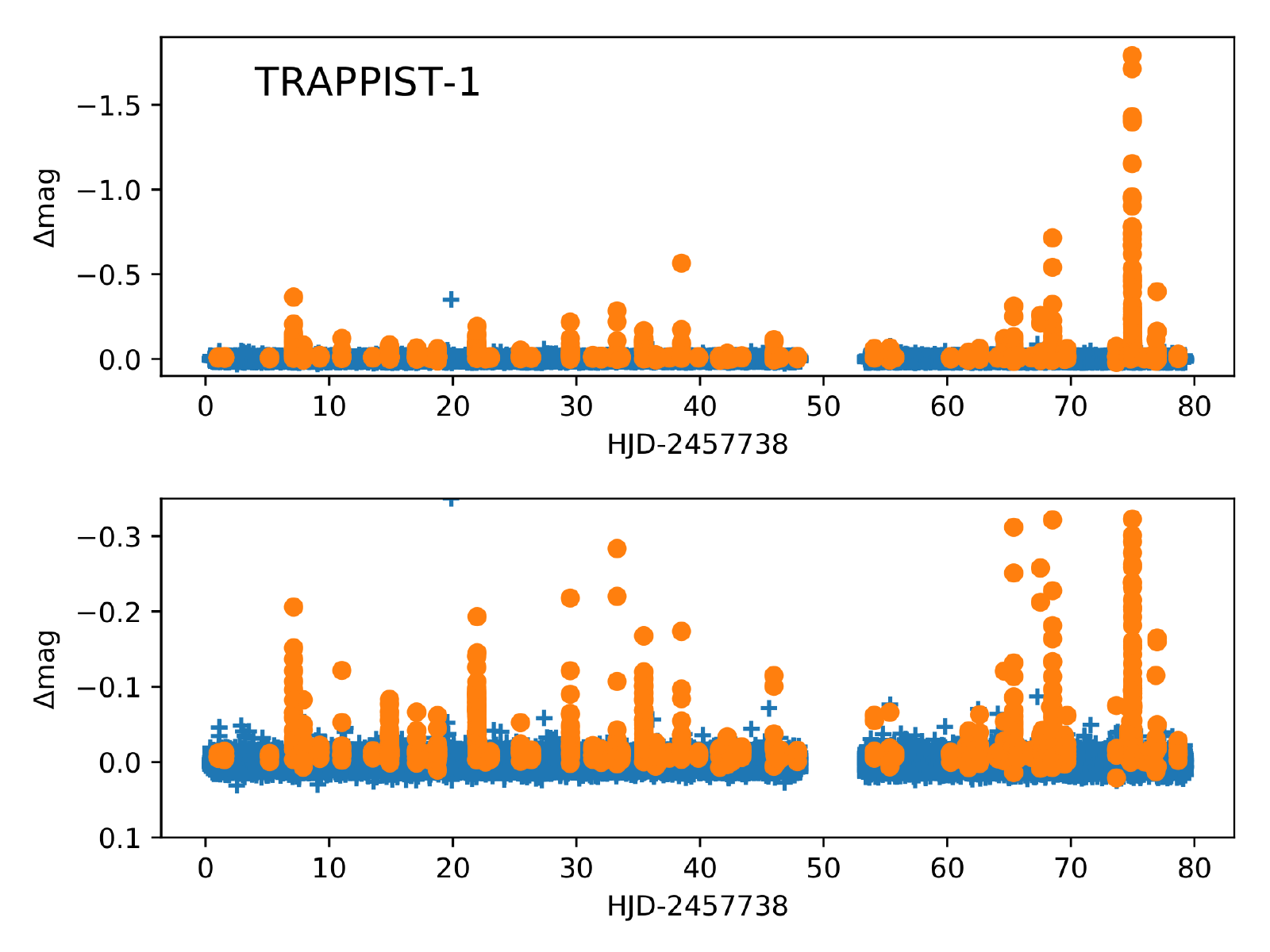}
\includegraphics[width=0.7\textwidth]{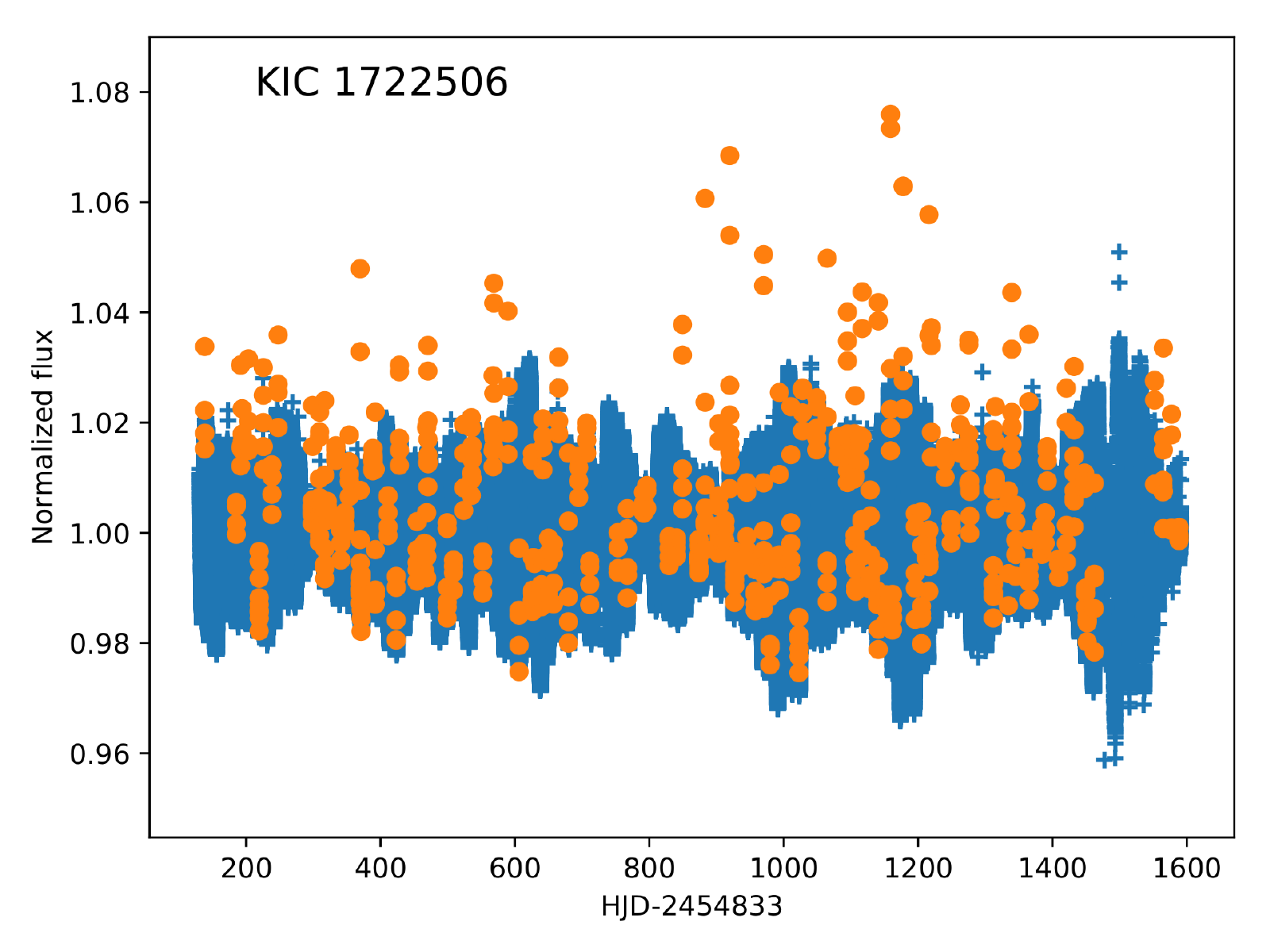}%
\caption{Top: Selected flare candidates in the TRAPPIST-1 short-cadence {\em K2} data. The upper panel shows the total light curve, the middle plot is zoomed-in to show smaller events. Bottom: Similar analysis, but for long-cadence Kepler data of KIC 1722506.}
\label{fig:results}
\end{center}
\end{figure*}
%%%%%%%%%%%%%%%%%%%%%%%%%%%%%%%%%%%%%%%%%%%%%%%%%%%%%
\section{Testing the \thecode{} code}   %  T E S T  % 
%%%%%%%%%%%%%%%%%%%%%%%%%%%%%%%%%%%%%%%%%%%%%%%%%%%%%

%%%%%%%%%%%%%%%%
\begin{figure*}
\begin{center}
\includegraphics[width=0.9\textwidth]{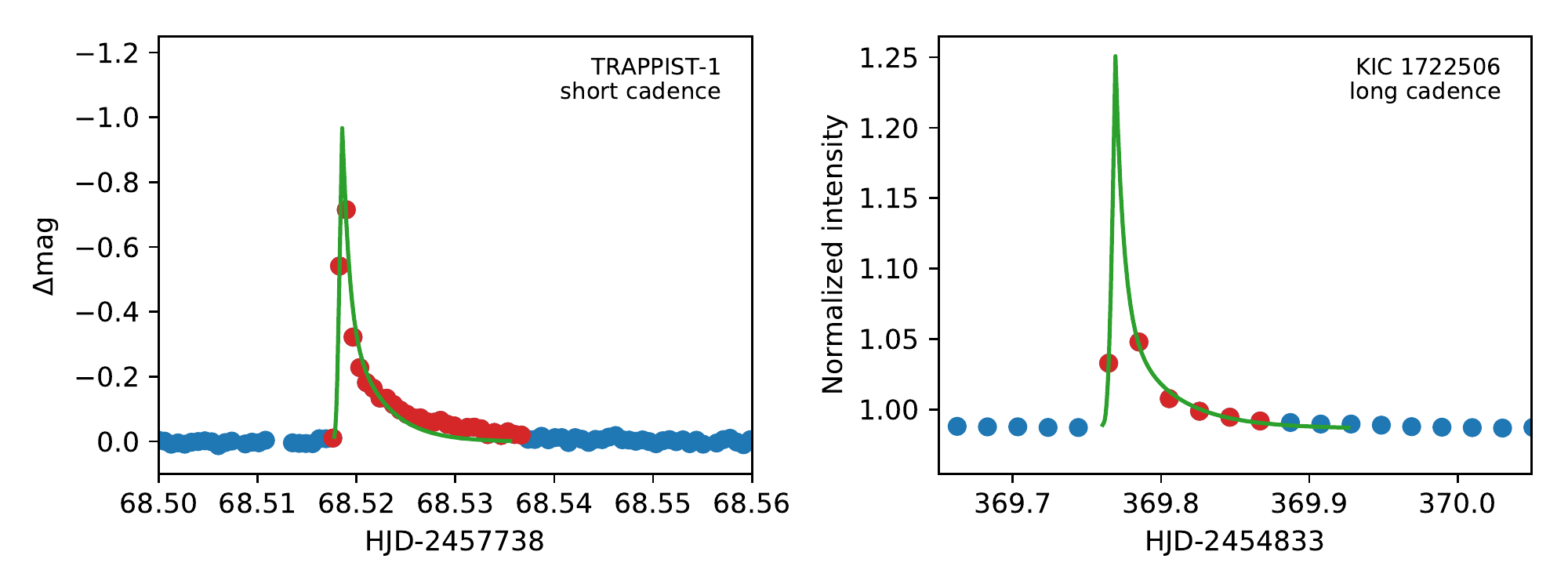}
\caption{Two samples from the recovered flares, from the short-cadence TRAPPIST-1 and the long-cadence KIC 1722506 data. Red points show the points selected for flares, the green line indicates the fitted analytical model from \cite{2014ApJ...797..122D}. }
\label{fig:samples}
\end{center}
\end{figure*}

As a demonstration of the \thecode{} code, we analyzed two data sets from the {\em Kepler} telescope: a short-cadence {\em K2} observation of TRAPPIST-1, and a long-cadence light curve of KIC 1722506 (the data were obtained in the 420--900\,nm wavelength range with the maximum spectral response at 575\,nm). In the case of TRAPPIST-1, we set the minimum number of data points needed for a flare (N3 in \citealt{2015ApJ...814...35C}) to 5, and the detection limit to $5\sigma$. In the case of KIC 1722506 we used N3=3 with a detection limit of $3\sigma$. The detected flare events are shown in Figure \ref{fig:results}.
Two samples from the recovered flares that demonstrate the analytic model fit by \thecode{} are plotted in Figure \ref{fig:samples}. 

 With these parameters, \thecode{} found 35 and 126 events  in the case of TRAPPIST-1 and KIC 1722506, respectively. As a comparison, in the case of TRAPPIST-1, visual inspection by \cite{trappist1} revealed 42 events.

To estimate the flare energies, we followed the method of \cite{flare-energy}, which is based on integrating the flare intensity during the event:
$$\varepsilon_f = \int\limits_{t_1}^{t_2} \left( \frac{I_{0+f}(t)}{I_0}-1\right)dt,$$
 where $t_1$ and $t_2$ are the beginning and end times of an event, and $I_{0+f}$ and $I_0$ are the intensities with and without a flare (i.e., their fraction is the normalized intensity). The integral above will yield
the relative flare energy (or equivalent duration). During the analysis, \thecode{} fits each flare event as described above and produces a spotless light curve that contains only the eruption itself, which has to be integrated over the duration of the event in order to obtain the relative flare energy. From this, the flare energy in the observed bandpass ($E_f$) can be calculated by multiplying by the quiescent flux ($F_\star$):
$$E_f = \varepsilon_f F_\star.$$
We estimated the quiescent flux by assuming blackbody radiation with an effective temperature of $T_\mathrm{eff}=2550$K and stellar radius of $R=0.117R_\odot$ for TRAPPIST-1 \citep{2016Natur.533..221G}, and $T_\mathrm{eff}=4270$K and $R=0.845R_\odot$ for KIC 1722506 \citep[taken from the {\em Kepler} Input Catalogue%
\footnote{\url{http://archive.stsci.edu/kepler/}, see also \citealt{KIC} };
][]{2011AJ....142..112B}. The blackbody power function $\mathcal{F}(\lambda)$ was convolved with the {\em Kepler} response function $S_\mathrm{Kp}$, and integrated over wavelength to obtain the quiescent flux $F_\star$ in the {\em Kepler} passband:
$$F_\star = 
\int\limits_{\lambda_1}^{\lambda_2} 4\pi R^2 \mathcal{F}(\lambda) S_\mathrm{Kp}(\lambda) d\lambda.$$

%%%%%%%%%%%%%%%%
\begin{figure*}
\begin{center}
\includegraphics[width=0.45\textwidth]{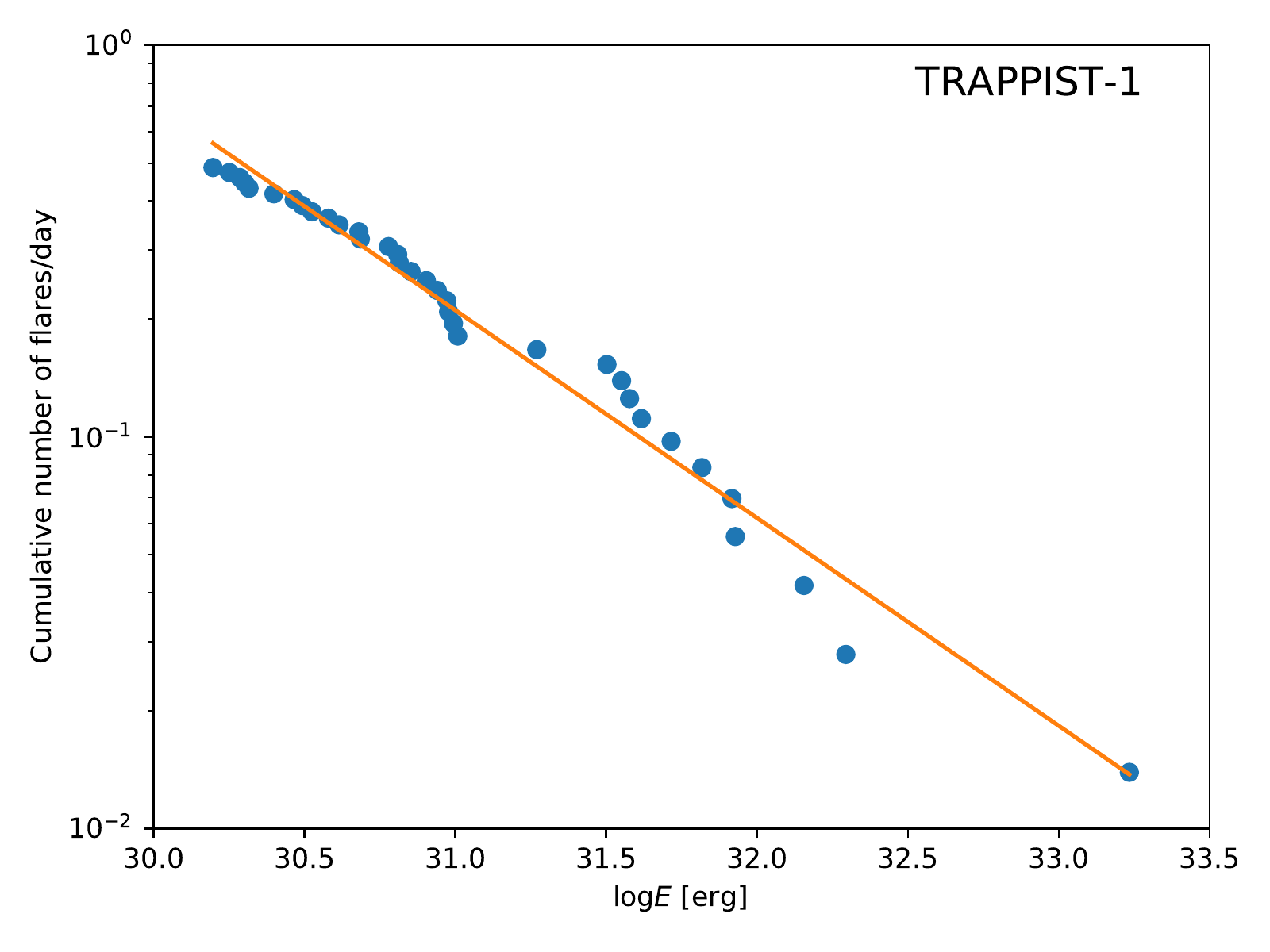}
\includegraphics[width=0.45\textwidth]{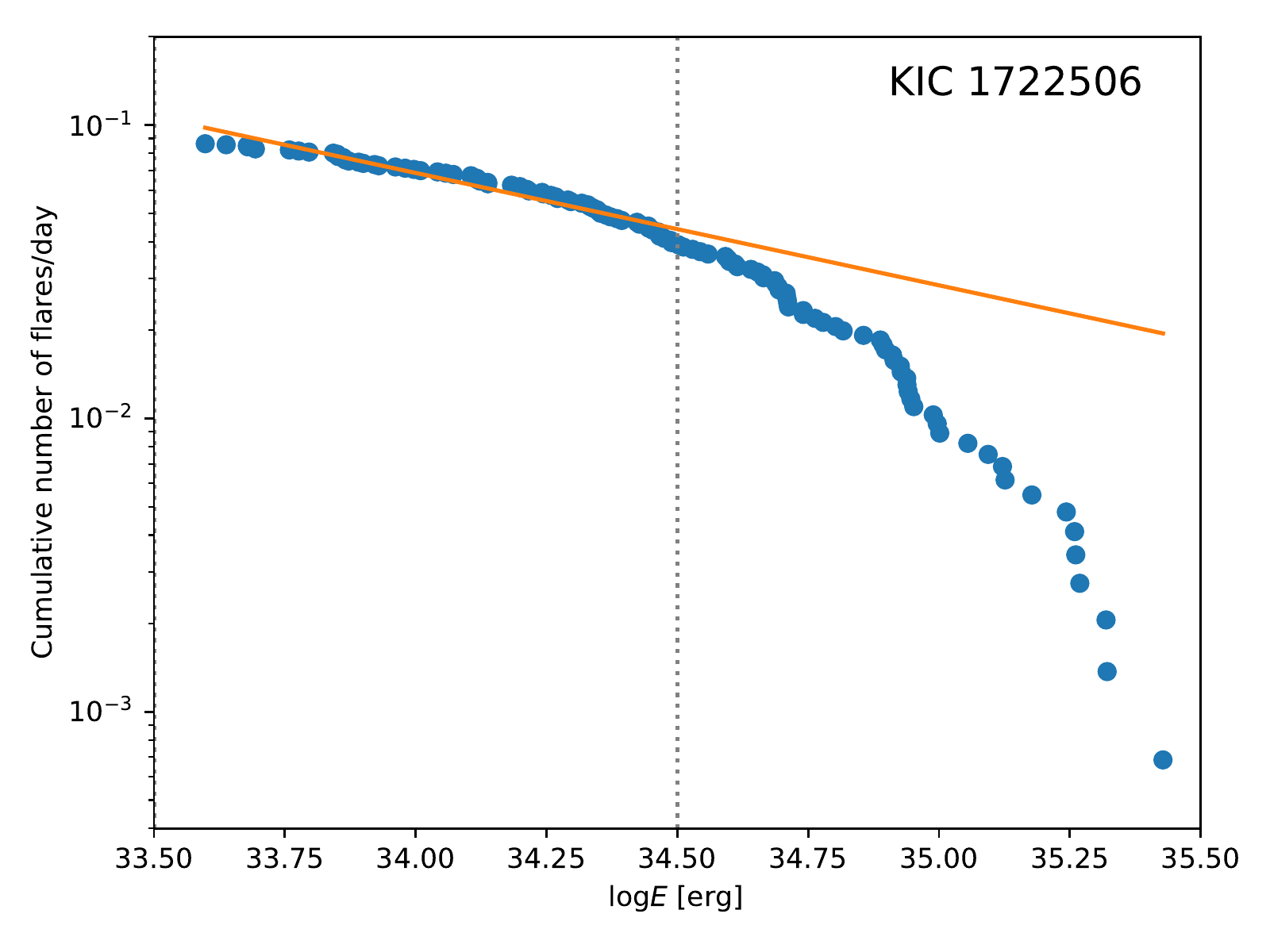}
\caption{Cumulative flare frequency distribution as a function flare energy, $\log E$ where $E$ is given in ergs, of the TRAPPIST-1 {\em K2} data (left, cf. \citealt{trappist1}) and for long-cadence Kepler data of KIC 1722506 (right). In the right plot, the dashed line marks the upper energy limit used for the fit.}. 
\label{fig:cumulative}
\end{center}
\end{figure*}

Following the analysis of \cite{trappist1}, we fit the cumulative flare frequency distribution 
(plotted in Fig. \ref{fig:cumulative}) with a linear fit that can be expressed as
$$\log \nu = a + \beta \log E,$$
where $\nu$ is the cumulative number of flares with energy higher than $E$. The slope of a linear fit yields $\beta = 1 - \alpha$, where $\alpha$ is often used to determine if flare energy dissipates by thermal, nonthermal, or magnetic processes (see \citealt{2016ApJ...832...27A}). Alternatively, $\alpha$ can be determined using the maximum likelihood estimator (see \citealt{maximum-likelihood}):
$$(\alpha -1) = n \left[\sum\limits_{i=1}^n \ln \frac{E_i}{E_\mathrm{min}}\right]^{-1}.$$
Here, $n$ is the number of detected events, while $E_i$ and $E_\mathrm{min}$ are the individual and the lowest flare energies, respectively. According to \cite{maximum-likelihood}, this result should be multiplied by $\frac{n-2}{n}$ for small samples to correct for the bias.
The linear fit to the cumulative distribution yielded $\alpha=1.53$ for TRAPPIST-1, while the maximum likelihood estimator gives $\alpha=1.47$ (corrected for sample size), close to $\alpha=1.59$, the value found by \cite{trappist1}.
For KIC 1722506, the linear fit yielded $\alpha=1.38$  for energies $\log E<34.5$ ($E$ given in ergs), where the distribution is close to linear, while the maximum likelihood estimator gave $\alpha=1.50$.

%%%%%%%%%%%%%%%%%%%%%%%%%%%%%%%%%%%%%%%%%
\section{Summary}       % S U M M A R Y %
%%%%%%%%%%%%%%%%%%%%%%%%%%%%%%%%%%%%%%%%%
We presented the \thecode{} code, which uses machine-learning methods to identify flare events in light curves and calculates their relative energies. Characterizing these energetic events is crucial, since they can shape circumstellar environments, especially in the realm of planetary habitability. In the case of many targets and large data sets, as with the {\em Kepler} database, manual inspection is impossible, but machine-learning tools can help astronomers to effectively analyze such data.
In the future, we plan to apply this method to a large set of {\em Kepler} stars in order to obtain a new view that is independent of currently available works.

\begin{acknowledgements}
The authors thank A. Mo\'or for useful discussion and the anonymous referee for their careful review of this work.
The authors acknowledge the Hungarian National Research, Development and Innovation Office
grants OTKA K-109276, OTKA K-113117, and supports through
the Lend\"ulet-2012 Program (LP2012-31) of the Hungarian Academy of Sciences,
and the ESA PECS Contract No. 4000110889/14/NL/NDe. 
KV is supported by the Bolyai J\'anos Research Scholarship of the Hungarian Academy of Sciences. 
This work has used {\em K2} data from the proposal number GO12046. Funding for the {\em Kepler} and {\em K2} missions is provided by the NASA Science Mission directorate. 
\end{acknowledgements}


\bibliographystyle{aa} 
\bibliography{paper}

\end{document}